\documentclass[apj]{emulateapj}

\def\simgt{\lower.5ex\hbox{$\; \buildrel > \over \sim \;$}}
\def\simlt{\lower.5ex\hbox{$\; \buildrel < \over \sim \;$}}

\def\etal{{et~al.}}
\def\amin{\ifmmode^{\prime}\else$^{\prime}$\fi}
\def\asec{\ifmmode^{\prime\prime}\else$^{\prime\prime}$\fi}

\def\simgt{\lower.5ex\hbox{$\; \buildrel > \over \sim \;$}}
\def\simlt{\lower.5ex\hbox{$\; \buildrel < \over \sim \;$}}

\newcommand\xte{{\it RXTE\/}}

\newcommand\einstein{{\it Einstein}}

\newcommand\sax{{\it Beppo}SAX\/}

\newcommand\asca{{\it ASCA\/}}

\newcommand\chandra{{\it Chandra}}

\newcommand\xmm{{\it XMM-Newton}}

\newcommand\hess{{H.E.S.S.}}

\newcommand\integral{{\it INTEGRAL\/}}

\newcommand\swift{{\it Swift\/}}

\newcommand\asrc{\hbox{AX~J1838.0$-$0655}}
\newcommand\bsrc{\hbox{AX~J1837.3$-$0652}}
\newcommand\tev{\hbox{HESS~J1837$-$069}}

\newcommand\psr{PSR~J1838$-$0655}

\slugcomment{To Appear in The Astrophysical Journal}

\shorttitle{Discovery of a Pulsar Associated with \tev}
\shortauthors{Gotthelf \& Halpern}

\begin{document}

\title{Discovery of a Young, Energetic 70.5 ms Pulsar Associated\\
with the TeV Gamma-ray Source \tev} 

\author{E. V. Gotthelf \&  J. P. Halpern}

\affil{Columbia Astrophysics Laboratory, Columbia University, 550 West
120$^{th}$ Street, New York, NY 10027, USA}

\begin{abstract}

We report the discovery of 70.5~ms pulsations from the X-ray source
\asrc\ using the {\it Rossi X-ray Timing Explorer} (\xte).
\psr\ is a rotation-powered pulsar with spin-down luminosity
$\dot E = 5.5\times 10^{36}$~ergs~s$^{-1}$, characteristic age
$\tau_c \equiv P/2\dot P = 23$~kyr, and surface dipole magnetic
field strength $B_s = 1.9\times10^{12}$~G.  
It coincides with an unresolved
\integral\ source and the extended TeV source \tev.  At an
assumed distance of 6.6~kpc by association with
an adjacent massive star cluster, the efficiency of \psr\
converting spin-down luminosity to radiation is 0.8\% for the
2$-$10~keV \asca\ flux, 9\% for the 20$-$300~keV \integral\
flux and $\sim 3\%$ for the $>200$ GeV emission of \tev,
making it a plausible power source for the latter.
A \chandra\ X-ray observation resolves \asrc\ into a bright
point source surrounded by a $\approx 2^{\prime}$ diameter,
centrally peaked nebula.
The spectra of the pulsar and nebula are each well fitted by
power laws, with photon indices $\Gamma = 0.5(0.3-0.7)$ and $\Gamma =
1.6(1.1-2.0)$, respectively. The 2$-$10~keV X-ray
luminosities of the pulsar and nebula are $L_{PSR} = 4.6
\times 10^{34}\ d_{6.6}^2$~ergs~s$^{-1}$ and $L_{PWN} = 5.2 \times
10^{33}\ d_{6.6}^2$~ergs~s$^{-1}$. 
A second X-ray source adjacent to the TeV emission,
\bsrc, is resolved into an apparent
pulsar/PWN; it may also contribute to \tev.
The star cluster RSGC1 may have given birth to one or both pulsars,
while fueling TeV emission from the extended PWN with target
photons for inverse Compton scattering.

\end{abstract}

\section{Introduction}

Surveys of the Galactic plane \citep{aha06} by the
High Energy Stereoscopic System (\hess) find
that at least half of its sources \citep{fun07}
can be identified with
supernova remnants (SNRs) or pulsar wind nebulae (PWNe).
The sizes of those sources that are
identified as PWNe are generally larger
in TeV $\gamma$-rays than in X-rays,
revealing a new view of the penetration of high-energy particles
into the surrounding medium.  The larger TeV nebulae are
often displaced from the pulsar, possibly by the reverse
shock of a supernova that exploded in an inhomogeneous medium.
They may also contain relic electrons
from the more energetic, younger phase of the pulsar spin-down.
The characteristic mismatch between X-ray and $\gamma$-ray
sizes, together with the positional offsets of
the X-ray and TeV emitting nebulae, are becoming familiar
as new examples are found \citep[][2006c, 2007]{aha06b}.
Understanding exactly
how the spin-down luminosity of a particular pulsar
powers its TeV nebula requires time-dependent modeling and
sometimes uncertain details of the local environment
that must be determined at other wavelengths.
The favored theoretical mechanisms, either inverse-Compton scattering
of ambient photons, or decay of neutral pions produced in hadronic collisions
of high-energy protons with a dense phase of the ISM, are being
tested in several cases.  For a  recent review of these issues, see
\citet{dej08}.

A most instructive association is \asrc\ with \tev.
One of the first extended
sources detected by \hess\ \citep{aha05,aha06}, it was considered
unidentified until now even though X-ray studies of the field pointed to a
coincident hard, steady X-ray source detected up to 300~keV by
\integral\ \citep{mal05}.
\asrc\ was seen by X-ray
satellites spanning decades, beginning with \einstein\
\citep[1E 1835.3$-$0658;][]{her88},
and including \asca\ \citep{bam03},
\sax\ \citep{mal05},
\xmm\ (unpublished),
and \swift\ \citep{lan06}, always with steady flux.
\citet{bam03} measured $1.1 \times 10^{-11}$ ergs cm$^{-2}$ s$^{-1}$
from \asrc\ in the 0.7$-$10 keV band, which is comparable to
the $> 200$~GeV flux from \tev, $3.3 \times 10^{-11}$
ergs~cm$^{-2}$~s$^{-1}$ \citep{aha06}.
The absence of variability is an important clue to the nature
of \asrc, as is its hard spectrum.  \citet{mal05} fitted the
combined \integral\ and \asca\ spectrum with a power law of
photon index $\Gamma=1.5$, and conjectured that the source is a supernova
product, more likely a PWN than a shell SNR.

\begin{figure*}[t]
\centerline{
\hfill
\includegraphics[height=0.475\linewidth,angle=270,clip=true]{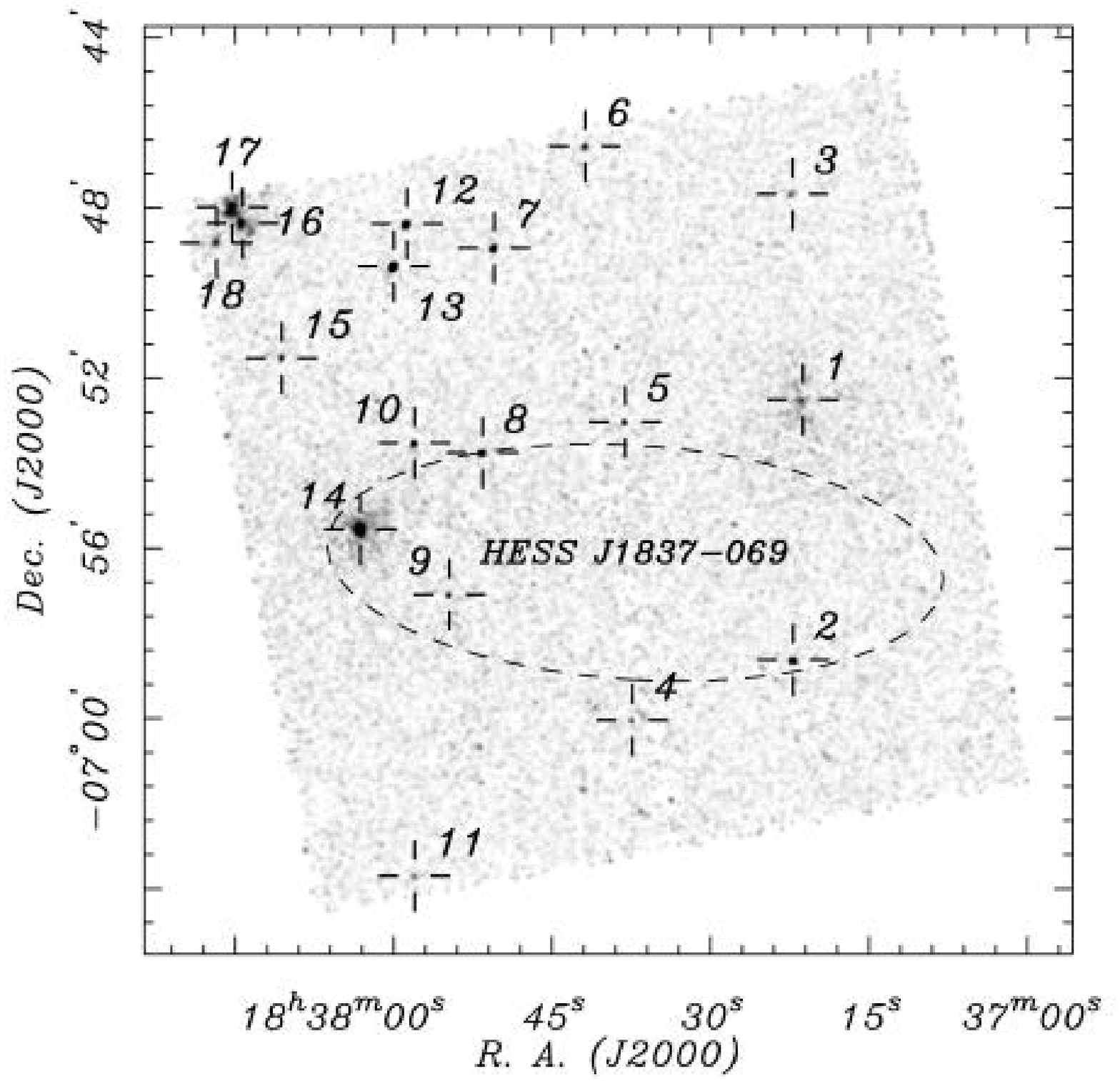}
\hfill
\includegraphics[height=0.475\linewidth,angle=270,clip=true]{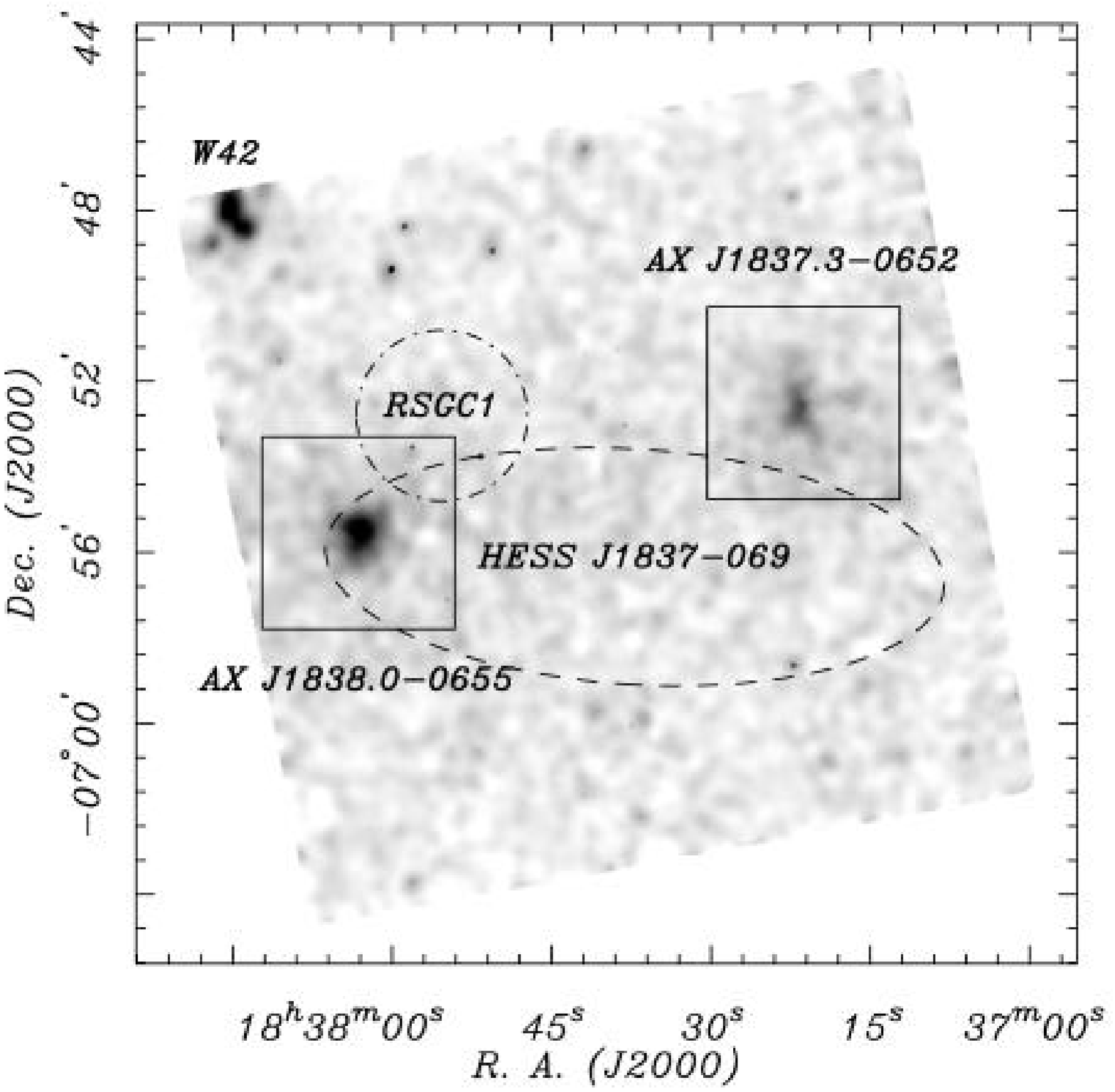}
\hfill
}
\caption{\chandra\ ACIS-I broad-band X-ray image of \tev.  {\it Left}:
Image scaled to emphasize point sources listed in Table~1.  {\it
Right}: Image smoothed to emphasize diffuse emission.  The ellipse
represents the two-dimensional Gaussian $1\sigma$ extent of the TeV
emission \citep{aha06}.  The circle is the approximate extent of the
red supergiant cluster RSGC1 \citep{fig06}. The squares indicate the
zoomed-in regions around \asrc\ and \bsrc\ corresponding to the insets
in Figure~\ref{fig:radial}.}
\label{fig:chandraimage}
\end{figure*}

In \S 2 we describe our analysis of
a \chandra\ image of \tev\ that resolves
\asrc\ into a pulsar/PWN, and study its spectral properties.
We also address the nature of the other
X-ray sources in the vicinity of \tev, one of which,
\bsrc, is evidently a second PWN.
Motivated by the \chandra\ results, we proposed observations
of \asrc\ with \xte\ to search for its putative pulsar.
The resulting discovery of 
\psr\ and measurement of its spin-down parameters
are described in \S 3.
We show that the rotational energy/spin-down luminosity of \psr\
are sufficient to power the TeV emission of \tev, 
and we further argue in \S 4 that the adjacent massive star cluster RSGC1
\citep{fig06,dav08} provides ambient target photons for inverse Compton
scattered TeV emission by PWN electrons from \asrc\ and
possibly from \bsrc, whose pulsar may also have been born in the cluster.

\begin{deluxetable*}{lccccccccl}[h]
\tabletypesize{\small}
\tablewidth{0pt}
\tablecaption{Point Sources in \chandra\ ACIS-I ObsID 6719\label{tab:chandratable}}
\tablehead
{
\colhead{No.} & \multicolumn{2}{c}{X-ray Position} & \colhead{Counts} & \colhead{HR\tablenotemark{a}} &
\multicolumn{2}{c}{Optical\tablenotemark{b} or Radio Position} & \colhead{$B2$\tablenotemark{b}} &
\colhead{$R2$\tablenotemark{b}} & \colhead{ID (Spectral Type)} \\
\colhead{} & \colhead{R.A. (J2000)} & \colhead{Decl. (J2000)} & \colhead{} & \colhead{} &
\colhead{R.A. (J2000)} & \colhead{Decl. (J2000)} & \colhead{(mag)} & \colhead{(mag)} & \colhead{}
}
\startdata
1\dotfill  & 18 37 21.21 & $-$06 52 30.6 & \phn\phn 17 & $+$0.77 & ... & ... & ... & ... & \bsrc \\
2\dotfill  & 18 37 22.14 & $-$06 58 37.5 & \phn\phn 83 & $-$0.36 & 18 37 22.126 & $-$06 58 38.09 & 16.73 & 14.30 & \hfil ... \hfil \\
3\dotfill  & 18 37 22.20 & $-$06 47 40.4 & \phn\phn 59 & $+$0.82 & ... & ... & ... & ... & \hfil ... \hfil \\
4\dotfill  & 18 37 37.37 & $-$07 00 02.6 & \phn\phn 19 & $-$0.58 & 18 37 37.374 & $-$07 00 02.47 & 16.50 & 13.64 & \hfil ... \hfil \\
5\dotfill  & 18 37 38.04 & $-$06 53 01.8 & \phn\phn 25 & $+$0.92 & ... & ... & ... & ... & \hfil ... \hfil \\
6\dotfill  & 18 37 41.84 & $-$06 46 33.0 & \phn\phn 45 & $+$0.86 & ... & ... & ... & ... & \hfil ... \hfil \\
7\dotfill  & 18 37 50.45 & $-$06 48 56.4 & \phn\phn 85 & $+$0.98 & 18 37 50.45  &  $-06$ 48 56.9 & ... & ... & GPSR5 25.320$-$0.098 \\
8\dotfill  & 18 37 51.57 & $-$06 53 45.4 & \phn    274 & $+$0.99 & 18 37 51.60  &  $-06$ 53 45.0 & ... & ... & GPSR5 25.252$-$0.139 \\
9\dotfill  & 18 37 54.71 & $-$06 57 05.6 & \phn\phn 23 & $+$0.91 & ... & ... & ... & ... & \hfil ... \hfil \\
10\dotfill & 18 37 58.00 & $-$06 53 31.6 & \phn\phn 55 & $+$0.96 & 18 37 57.99  &  $-06$ 53 31.0 & ... & ... & GPSR5 25.266$-$0.161 \\
11\dotfill & 18 37 58.00 & $-$07 03 42.4 & \phn\phn 44 & $+$0.85 & ... & ... & ... & ... & \hfil ... \hfil \\
12\dotfill & 18 37 58.72 & $-$06 48 22.5 & \phn    129 & $-$0.87 & 18 37 58.760 & $-$06 48 22.39 & \phn 8.76 & \phn 7.75 & HD 171999 (G5) \\
13\dotfill & 18 37 59.97 & $-$06 49 22.4 & \phn    285 & $+$0.96 & ... & ... & ... & ... & Swift J1837.9$-$0649 \\
14\dotfill & 18 38 03.13 & $-$06 55 33.4 &        2114 & $+$0.92 & ... & ... & ... & ... & \asrc \\
15\dotfill & 18 38 10.59 & $-$06 51 31.9 & \phn\phn 46 & $+$0.80 & ... & ... & ... & ... & \hfil ... \hfil \\
16\dotfill & 18 38 14.26 & $-$06 48 21.7 & \phn    240 & $+$0.65 & ... & ... & ... & ... & \hfil ... \hfil \\
17\dotfill & 18 38 15.27 & $-$06 47 58.9 & \phn    291 & $+$0.42 & 18 38 15.231 & $-$06 47 58.78 &  ...  & 16.06 & W42 1 (O5.5) \\
18\dotfill & 18 38 16.75 & $-$06 48 49.4 & \phn    126 & $-$0.06 & ... & ... & ... & ... & \hfil ... \hfil
\enddata
\tablecomments{\footnotesize Includes all point sources that have signal-to-noise ratio $>3$
in the 0.3$-10$~keV band, as determined by the CIAO software package
source-detection tool {\tt wavdetect}.
Units of right ascension are hours, minutes, and
seconds, and units of declination are degrees, arcminutes, and arcseconds.}
\tablenotetext{a}{\footnotesize Hardness ratio defined as HR = $(N_h-N_s)/(N_h+N_s)$,
where $N_s$ and $N_h$ are the counts measured in the
$0.3-2$~keV and $2-10$~keV energy band, respectively.}
\tablenotetext{b}{\footnotesize Optical data from the USNO-B1.0 catalog (Monet et al. 2003).}
\end{deluxetable*}

\begin{figure*}[t]
\hspace{-0.1in}
\centerline{
\hfill
\includegraphics[height=0.38\linewidth,angle=0,clip=true]{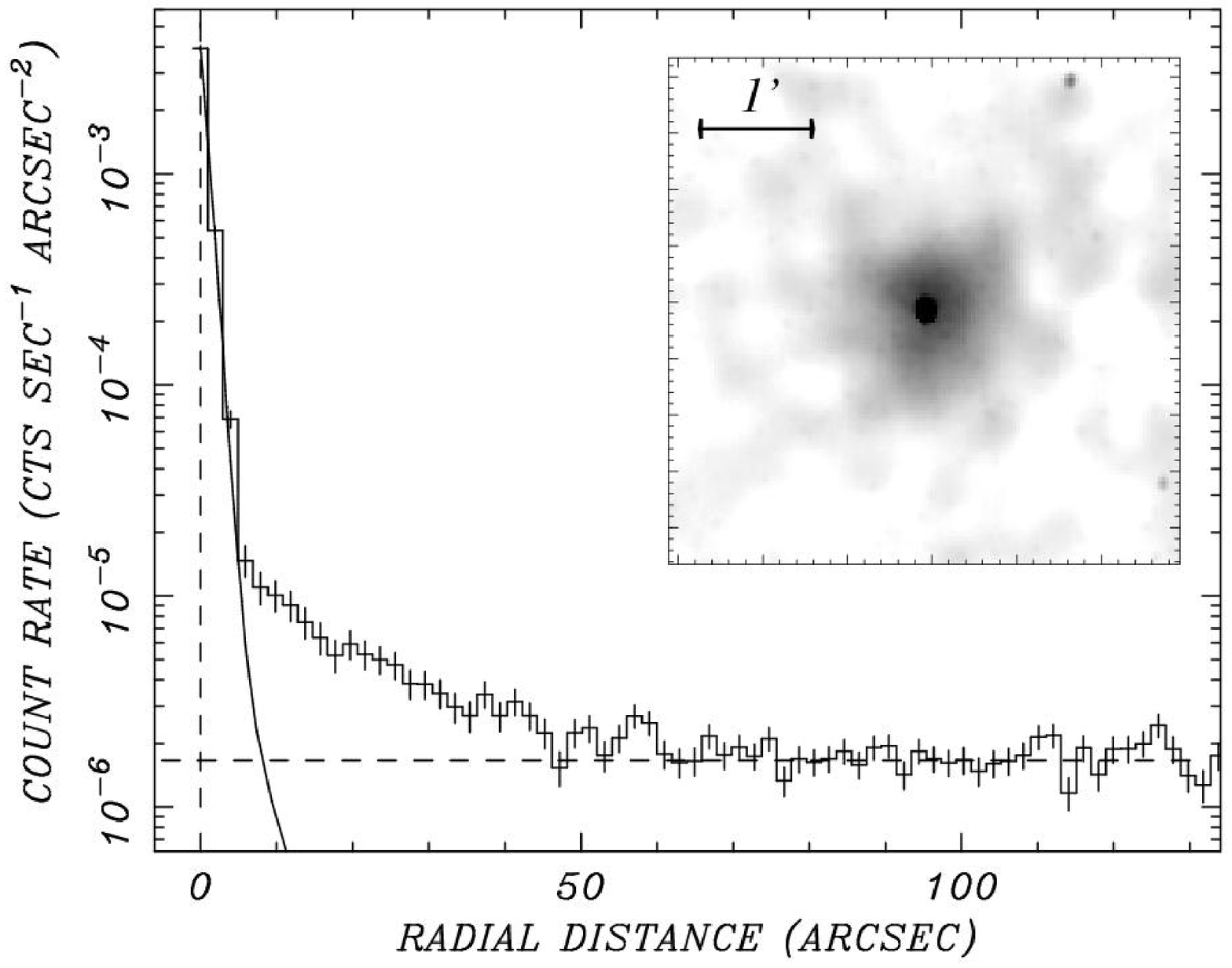}
\hfill
\includegraphics[height=0.38\linewidth,angle=0,clip=true]{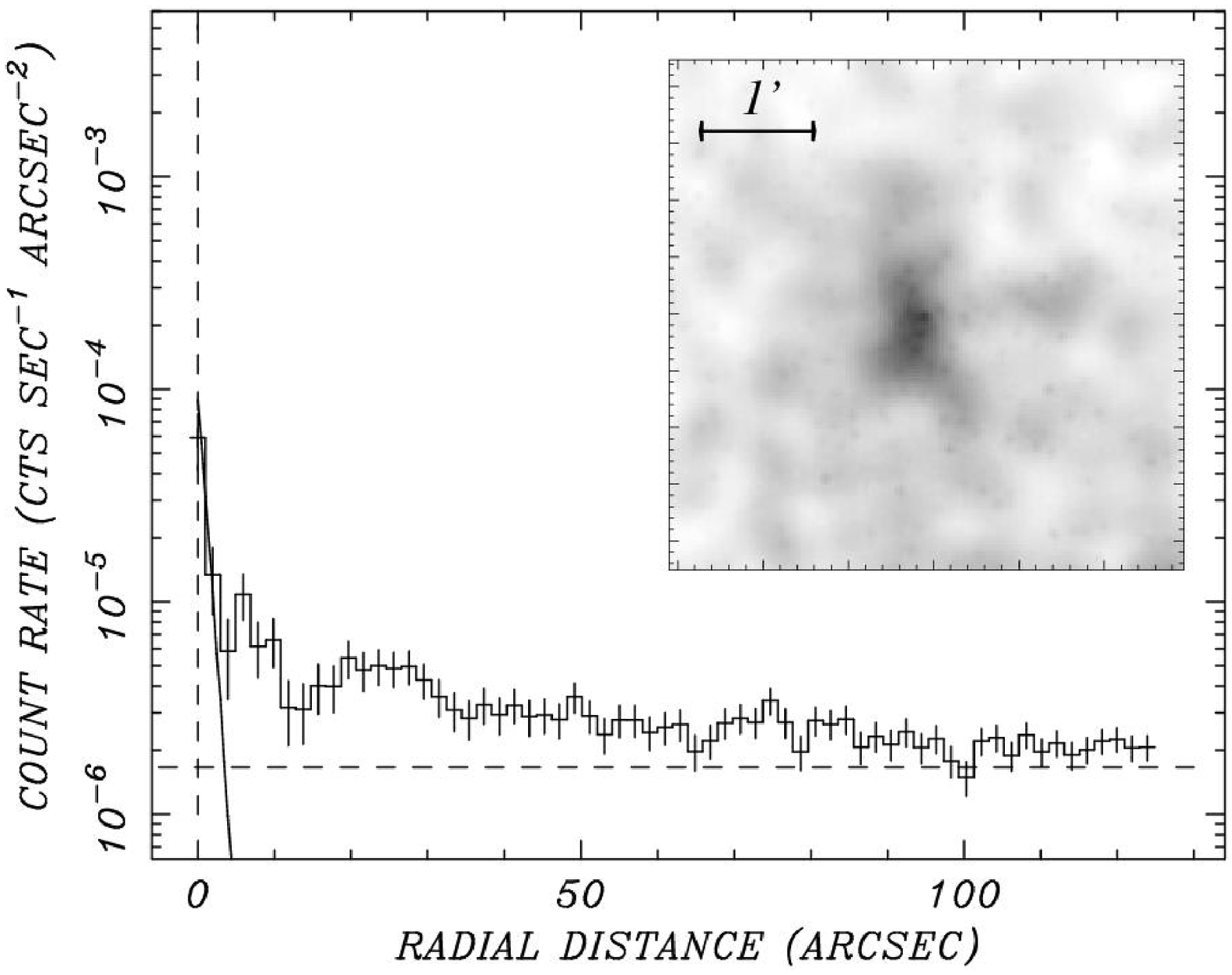}
\hfill
}
\caption{
\chandra\ ACIS radial
profiles of \asrc\ ({\it left\/}) and \bsrc\ ({\it right})
compared to the scaled point spread function at their locations
(FWHM $\approx 2^{\prime\prime}$;
{\it solid line}).  Surrounding each point source, a clear flux excess
is evident over the field background flux level
({\it horizontal dashed line}) to $r \approx 60^{\prime\prime}$
in the case of \asrc.  For \bsrc, a larger but
lower surface brightness extension is seen.
{\it Insets}: Zoom-in on the \chandra\ images. The
intensity is scaled to highlight the diffuse emission;
the point sources are saturated. }
\label{fig:radial}
\end{figure*}

\section{{\it Chandra\/} Observations and Results}

A 20~ks \chandra\ X-ray observation of \tev\ was made on 2006 August 19
UT using the Advanced CCD Imaging Spectrometer \citep[ACIS;][]{bur97}
operating in the full-frame TIMED/VFAINT exposure mode. This
detector is sensitive to X-rays in the 0.3$-$12.0~keV energy range with
a resolution of $\Delta E /E \sim 0.06$ FWHM at 1~keV. The imaging
system offers an on-axis spatial resolution of $\approx 0.5^{\prime\prime}$, which
is also the instrument pixel size, and increases to $\sim 5^{\prime\prime}$ at
$10^{\prime}$ off-axis. A total of 19.9~ks of
live-time was accumulated with a CCD frame time of 3.241~s (given the
1.3\% readout deadtime).  With a maximum count rate in a pixel of $<
0.004$~s$^{-1}$, photon pile-up can safely be ignored. 
As normal procedure,
the spacecraft was dithered to average out the single-pixel spectral
response.  We used the
standard processed and filtered event data.
No time filtering was necessary as the background rate was
stable over the course of the observation.
All data reduction and analysis was performed using the CIAO (V3.4),
FOOLS (V6.0.4), CALDB 3.4.2, and XSPEC (V12.2.1) X-ray analysis
software packages.  We followed the CIAO online science thread to
create the exposure-corrected image.

Figure~\ref{fig:chandraimage}
is the exposure-corrected map of the entire 4-CCD ACIS-I
field that is smoothed on two scales to
highlight either the point sources or the diffuse emission.
There are 18 unambiguous point sources
(listed in Table~\ref{tab:chandratable}); two of these, sources 1 and 14,
have associated X-ray nebulae. 
Since the image was centered on the TeV emission,
\asrc\ (source 14) fell $5.25^{\prime}$ off-axis, near
the gap between ACIS-I CCDs. Fortunately, exposure
correction using the spacecraft dithering information fully recovers
the fluxed image, allowing us to generate an accurate spectrum and
radial profile.

\subsection{\asrc }

Figure~\ref{fig:chandraimage} shows that
\asrc\ falls along the long
axis of the TeV extent\footnote {Figure~6 of
\citet{lan06} that graphs the TeV extent on the \swift\ X-ray
image has the wrong position angle for the ellipse, which doesn't do
justice to the case for association.  We reproduce here the
orientation given by \citet{aha06}.}. Considering the off-axis
location of the point source, we determine its precise location by fitting
the off-axis point spread function following the CIAO recipe.
Then, by registering other
X-ray point sources with their optical or radio counterparts
(see Table~1) we are able to refine the location of
the point source in \asrc\ to (J2000.0)
R.A. $= 18^{\rm h}38^{\rm m}03.13^{\rm s}$,
decl. $ = -06^{\circ}55^{\prime}33.4^{\prime\prime}$
with a $1\sigma$ error radius of $\approx 0.3^{\prime\prime}$.
(A correction of $-0.55^{\prime\prime}$ in R.A. 
and $-0.09^{\prime\prime}$ in decl. was applied.)

Figure~\ref{fig:radial}
shows an enlarged region of the \chandra\ image of \asrc, and
its radial profile, clearly
resolving a nebula from a point source.
For spectral analysis of \asrc\ we extract photons in the
2$-$10~keV energy band, below which the spectra are highly attenuated
by the large column density.
For the point-source spectrum, we extract 2124
photons from a $5^{\prime\prime}\times7^{\prime\prime}$
elliptical aperture centered
on the source peak; the diffuse emission contributes a negligible
background (21 counts) in this region. For the nebula we use a
$1^{\prime}$ radius extraction region, offset from the point source
and centered at (J2000.0)
R.A. = $18^{\rm h}38^{\rm m}03.1^{\rm s}$, decl. =
$-06^{\circ}55^{\prime}40^{\prime\prime}$ with a
$7^{\prime\prime}\times9^{\prime\prime}$ region around the pulsar
removed.  The background for the nebula was extracted from a concentric
annulus of radii $ 1^{\prime} < r \leq 2^{\prime}$.
The nebula so extracted comprises 483 photons after subtracting
background (33\%) and excising the point source.

\begin{figure}[t]
\centerline{
\includegraphics[height=0.95\linewidth,angle=270,clip=true]{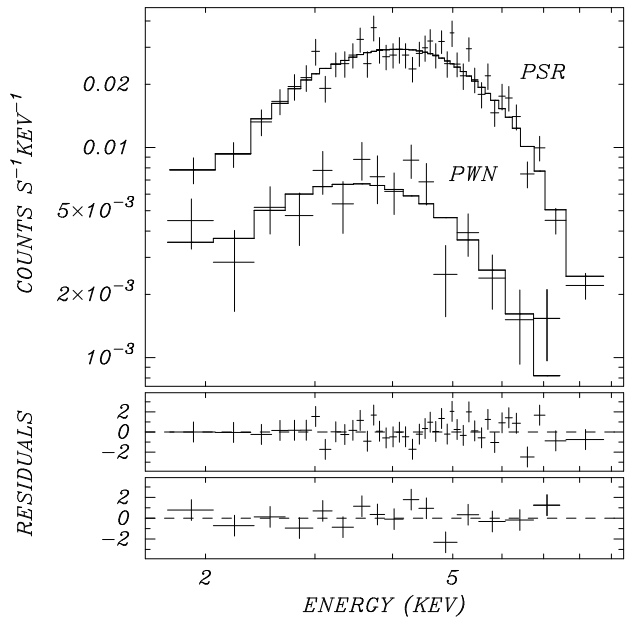}
}
\caption{\chandra\ ACIS spectra of \psr\ and its nebula, 
comprising \asrc, both fitted to
an absorbed power-law model (see text).
Residuals for each spectrum are shown in
units of standard deviation.}
\label{fig:chandraspectra}
\end{figure}

Spectra from the point source and nebula region were grouped with a
minimum of 50 counts per spectral channel and fitted using XSPEC.
The two spectra were fitted
simultaneously using absorbed non-thermal power laws with
their hydrogen column density parameters linked; the results are
summarized in Figure~\ref{fig:chandraspectra}.
The chosen model produced excellent
fits ($\chi^2=0.9$ for 57 DoF) for a common column density of $N_{\rm
H} \approx 4.5(3.7-5.2) \times 10^{22}$~cm$^{-2}$ (90\% confidence
interval is used throughout). The spectrum of the putative pulsar
has a photon index of $\Gamma = 0.5(0.3-0.7)$, similar to that
found for the pulsar PSR~J1811$-$1926 in the young SNR~G11.2$-$0.3, and at
the low (hard) end of that expected for a high-$\dot E$ rotation-powered
pulsar ($> 4\times 10^{36}$~erg~s$^{-1}$; cf. Gotthelf 2003).  For the
PWN emission we measure a photon index of $\Gamma = 1.6(1.1-2.0)$,
which is also at the low end for a PWN.
The unabsorbed 2$-$10~keV fluxes for the
putative PSR and PWN are $F_{PSR} = 8.8 \times
10^{-12}$~ergs~cm$^{-2}$~s$^{-1}$ and $F_{PWN} = 1.0 \times
10^{-12}$~ergs~cm$^{-2}$~s$^{-1}$, respectively. These results are
consistent with the \asca\ composite
spectrum, having photon index $\Gamma = 0.8$,
$N_{\rm H} = 4 \times 10^{22}$~cm$^{-2}$, and
$F(0.7-10\ {\rm keV}) = 1.1 \times 10^{-11}$ ergs cm$^{-2}$ s$^{-1}$
\citep{bam03}.  The flux ratio $F_{PWN}/F_{PSR} = 0.11$ is at the
low end of the distribution of PWNe observed by \chandra,
which range from 0.1$-$10 as compiled by \citet{kar08}.

\begin{figure}[b]
\centerline{
\includegraphics[height=0.95\linewidth,angle=0,clip=true]{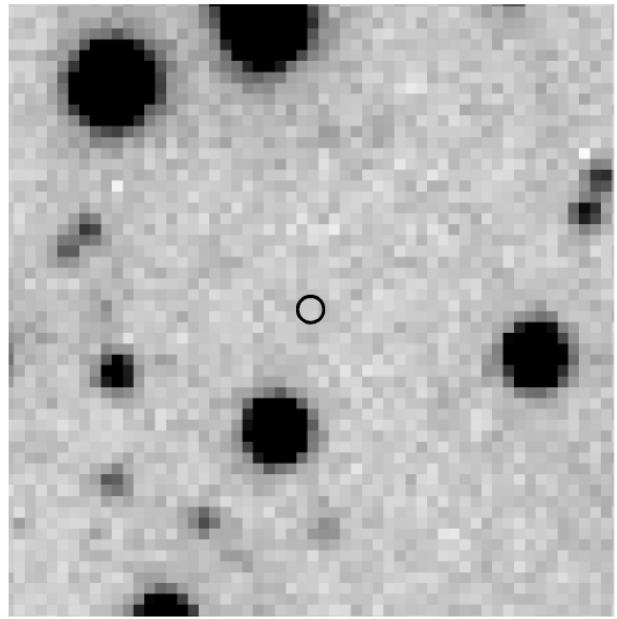}
}
\caption{A 600~s $R$-band exposure of the center of \asrc\
obtained on the 2.4m Hiltner telescope
on 2005 July 3.  The displayed field is
$30^{\prime\prime}\times30^{\prime\prime}$.
North is up, and east is to the left.
The \chandra\ position of \psr\ is indicated by
a circle of radius $0.7^{\prime\prime}$, approximately its
2 $\sigma$ uncertainty. The magnitude limit is $R=23.0$.}
\label{fig:mdmimage}
\end{figure}

The \chandra\ position of the point source
rules out the optical/IR candidate suggested
by \citet{lan06} on the basis of a less accurate \swift\ position.
We determine an optical upper limit of $R>23.0$
from an image obtained on the 2.4m Hiltner telescope
of the MDM Observatory (Fig.~\ref{fig:mdmimage}).
One does not expect to detect a pulsar having
the large distance and extinction implied by
the fitted $N_{\rm H}$ in an optical image of this depth.

\subsection{\bsrc}

In addition to \asrc, a fainter point source
(number 1 in Table~\ref{tab:chandratable})
of 17 photons is surrounded by a large, diffuse nebula.
It is evidently also a pulsar/PWN system.  This source was
detected by \asca\ (\bsrc; Bamba et al. 2003), with a 0.7$-$10~keV
flux of $1.8 \times 10^{-12}$ ergs~cm$^{-2}$ s$^{-1}$,
and by \swift\ (sources 6 and 7 of Landi et al. 2006).
It is difficult to measure here due to its low
surface brightness.  It is more extended than \asrc\
(see Figs.~\ref{fig:chandraimage} and \ref{fig:radial})
but has comparable total nebular flux,
$\approx 1.2 \times 10^{-12}$~ergs~cm$^{-2}$~s$^{-1}$
(unabsorbed) in the 2$-$10~keV band.  Unlike \asrc,
in \bsrc\ the PWN flux dominates over the
(putative) pulsar point source.
Its spectral parameters are largely
unconstrained, with $0.7 < \Gamma < 3.6$
and $N_{\rm H} = (2-12)\times 10^{22}$~cm$^{-2}$.
If \asrc\ is to be advanced as the origin of \tev,
then it cannot be ruled out that the PWN \bsrc\ 
makes a significant contribution.

\subsection{Additional \chandra\ X-ray Sources}

Sources 16$-$18 of Table~\ref{tab:chandratable}
lie in the northeast corner of the
\chandra\ image, and are evidently associated with the W42 star forming
region, with source 17 being the central O5.5 star in the cluster
\citep{blu00}.
Three additional sources (numbers 2, 4, and 12) are also stars, which is 
consistent with their softer X-ray spectra.  HD 171999 
in particular is a high-proper-motion
star.  Its optical position listed in Table~\ref{tab:chandratable}
is calculated for the epoch of the \chandra\
observation.  None of the other sources have optical/IR counterparts
in the Digitized Sky Survey or 2MASS; they are probably active
galactic nuclei.

Positions for radio sources denoted by GPSR5 designations in Table~1
are measured from the 1.4 GHz VLA Multi-Array Galactic Plane Survey
\citep{whi05,hel06}\footnote{See http://third.ucllnl.org/gps/.}.
The two radio sources nearest the TeV emission (numbers 8 and 10) were
discussed by \citet{fig06} and \citet{tre06}.  The latter
authors proved that they are extragalactic based on the kinematics
of their \ion{H}{1} absorption spectra.  Their X-ray positions
coincide with compact components of the radio sources, presumably
their host galaxy nuclei or quasars.  We have no reason
to hypothesize that any of these sources are related to \tev.

\section{{\it RXTE\/} Observations and Results}

The point-source flux measured by \chandra\ in \asrc\ is sufficient
that a pulsar search with \xte\ \citep{bra93} was deemed feasible.
Accordingly, we requested six observations in total, pointed at \asrc,
that spanned 2008 February 17 - March 5 UT to search for the expected
pulsar signal and develop a phase-connected timing solution.  A log of
the observations is given in Table~\ref{tab:xtelog}.  The data used
here were collected with the Proportional Counter Array
\citep[PCA;][]{jah96} in the GoodXenon mode with an average of 3.6 out
of the five proportional counter units (PCUs) active. In this mode,
photons are time-tagged to $0.9$ $\mu$s and have an absolute
uncertainty better than 100 $\mu$s \citep{rot98}. The effective area
of five combined detectors is about $6500$ cm$^{2}$ at 10~keV with a
roughly circular field-of-view of $\sim 1^{\circ}$ FWHM. Spectral
information is available in the 2$-$60~keV energy band with a
resolution of $\sim 16\%$ at 6~keV.

\begin{deluxetable}{llclc}[t]
\tabletypesize{\small}
\tablewidth{0pt}
\tablecaption{\label{tab:xtelog}Log of RXTE Observations and Period Measurements}
\tablehead{
\colhead{Date} & \colhead{Exp./Span} & \colhead{Start Epoch} & \colhead{Period\tablenotemark{a}} & \colhead{$Z^2_1$} \\
\colhead{(UT)} & \colhead{(ks)} & \colhead{(MJD)}                    & \colhead{(ms)}                    & \colhead{}
}
\startdata
2008 Feb 17           & \phantom{2}9.1/14.0 & 54513.825  & \phantom{8}70.49820(3) & \phantom{1}277\\
2008 Feb 19           &           20.1/35.0 & 54515.476  & \phantom{8}70.498218(6)& \phantom{1}726\\
2008 Feb 22           & \phantom{2}5.4/8.0  & 54518.609  & \phantom{8}70.49828(8) & \phantom{1}153\\
2008 Feb 26           & \phantom{2}5.5/8.2  & 54522.004  & \phantom{8}70.49823(6) & \phantom{1}186\\
2008 Mar \phantom{1}2 & \phantom{2}5.3/8.0  & 54527.378  & \phantom{8}70.49829(6) & \phantom{1}223\\
2008 Mar \phantom{1}5 & \phantom{2}6.3/8.6  & 54530.185  & \phantom{8}70.49830(6) & \phantom{1}206
\enddata
\tablenotetext{a}{\footnotesize Period derived from a $Z^2_1$ test. The \citet{lea83} uncertainty on the last digit is in parentheses.}
\end{deluxetable}

Standard time filters were applied to the realtime PCA data, which
rejects intervals of South Atlantic Anomaly passages, Earth
occultations, and other periods of high particle activity.  Our data
contained several short ($\sim 200$~s) intervals of detector arcing
which were excised. The realtime data required manually adjusting the
good data intervals to properly correspond to actual data collection
times for each PCU.  Table~\ref{tab:xtelog} lists the net exposure times
and total spans of
each data set. The photon arrival times were transformed to the solar
system barycenter in Barycentric Dynamical Time (TDB)
using the JPL DE200 ephemeris and the \chandra\
measured coordinates.

\subsection{Timing Analysis}

We restricted the analyzed data to the
2$-$20~keV energy range (PCA channels 2$-$50) from the top Xenon layer
of each PCU to optimize the signal-to-noise.  A Fast Fourier Transform
(FFT) of the first observation (February 17) revealed a significant
signal of power $S = 117$ at a period of $P=70.498$~ms\footnote{While planning these
observations, the signal was first found by C. Markwardt
(personal communication) in summed FFTs of hundreds of short, serendipitous
\xte\ dwells on the source taken during 2004-2008. However, the individual
$\sim 100$~s exposures did not contain sufficient statistics to
accurately measure the period or derive an ephemeris.}.  For the
$5.7 \times 10^6$ search elements this corresponds to essentially nil
false detection probability.  There is no doubt that \psr\ is the
compact object in \asrc, given the morphological evidence from \chandra\
and because \asrc\ is the brightest known X-ray source in the \xte\
field of view.

\begin{figure}
\centerline{
\includegraphics[height=0.95\linewidth,angle=270,clip=true]{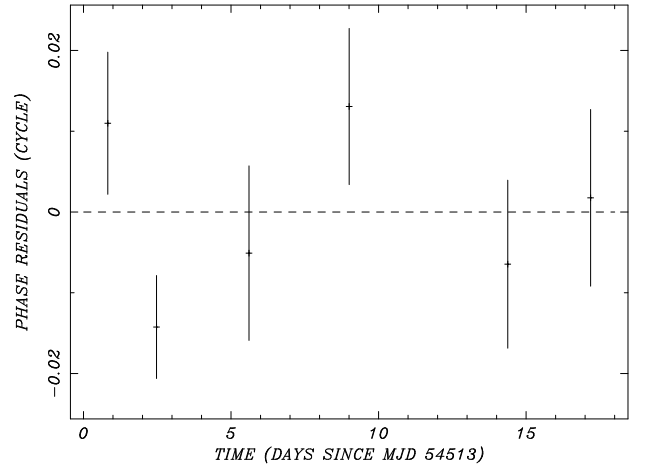}}
\caption{
Phase residuals of the six observations of \psr\ from
the quadratic ephemeris of Table~\ref{tab:ephem}.
}
\label{fig:residuals}
\end{figure}

Based on the strength of the signal in the February 17 and 19 detections
of \psr, the next four observations were scheduled from February~22 -
March~5 to develop a fully phase-coherent timing solution including
the previous data sets. For each epoch, we extracted the pulse profile
corresponding to the peak period as determined by the
$Z^2_1$ test \citep{buc83}. The resulting profiles were cross correlated,
shifted, and summed to generate a master pulse profile template.
Individual profiles were then
cross correlated with the template to determine the time of arrival
(TOA) and its uncertainty in phase at each epoch.  These TOAs were
iteratively fitted to a quadratic ephemeris using the TEMPO
software. We started by fitting just TOAs from the first three epochs,
which straddle the longest observation (February 19), to a linear
solution, and then added the last three TOAs to the fit one at a
time. At each step we found that the new TOA would match to $<0.1$
cycles the predicted phase derived from the previous set.
The complete resulting ephemeris is presented in Table~\ref{tab:ephem},
and the phase residuals are shown in Figure~\ref{fig:residuals}.
The residuals are all less than 0.02 cycles and appear to be random
within their statistical uncertainties.  Thus, inclusion of higher order
terms is not indicated.
An independent test of the uniqueness of the timing solution
was performed using a $Z_1^2$ search on a two-dimensional grid of $P$
and $\dot P$, from which we
obtained consistent ephemeris parameters and errors.

\begin{deluxetable}{ll}
\tabletypesize{\small}
\tablecaption{\label{tab:ephem}Timing Parameters of \psr }
\tablehead{
\colhead{Parameter}   &
\colhead{Value}   }
\startdata                                       
R.A. (J2000)\tablenotemark{a}\dotfill          & $18^{\rm h}38^{\rm m}03.13^{\rm s}$\\
Decl. (J2000)\tablenotemark{a}\dotfill         & $-06\arcdeg55'33.4^{\prime\prime}$    \\
Epoch (MJD TDB)\dotfill                        & 54522.00000012                \\
Period, $P$ (ms)\dotfill                       & 70.498243969(54)              \\
Period derivative, $\dot P$\dotfill            & $4.925(29)\times10^{-14}$       \\
Range of timing solution (MJD)\dotfill         & 54513--54530       \\
Characteristic age, $\tau_c$ (kyr)\dotfill      & 22.7                       \\
Spin-down luminosity, $\dot E$ (ergs\,s$^{-1}$)\dotfill & $5.5\times10^{36}$   \\
Surface dipole magnetic field, $B_s$ (G)\dotfill & $1.9\times10^{12}$
\enddata
\tablecomments{\footnotesize TEMPO $1\sigma$ uncertainties given in parentheses.}
\tablenotetext{a}{\footnotesize Chandra ACIS-I position from Table~\ref{tab:chandratable}.}
\end{deluxetable}

\begin{figure}[t]
\hspace{0.1in}
\includegraphics[height=0.9\linewidth,angle=270,clip=true]{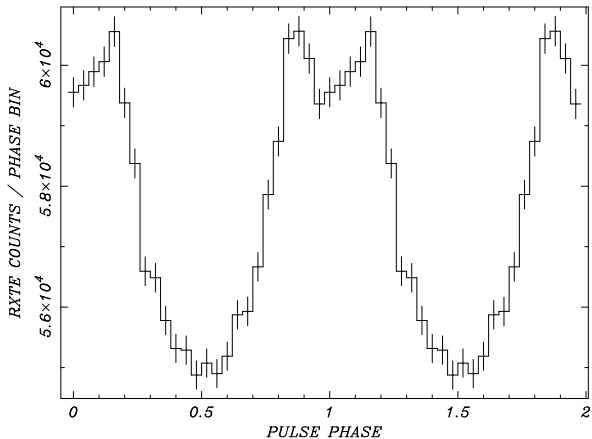}
\caption{\xte\ folded light curve of \psr\ in the 2$-$20 keV band.
Phase zero corresponds to the epoch of the ephemeris in Table~\ref{tab:ephem}.}
\label{fig:xtepulse}
\end{figure}

Figure~\ref{fig:xtepulse} displays the summed pulse profile using 
all the 2$-$20 keV data folded on the final ephemeris.  It has a symmetric,
blended double-peaked structure.
The measured pulsed fraction of $4.8\%$ indicates a
intrinsic modulation of $\simgt 50\%$ after allowing for the estimated
internal and Galactic background in the PCA.  
We see no energy dependence of the pulse profile
when subdividing the 2$-$20~keV band.

The spin-down parameters of \psr,
particularly its $\dot E = 5.5 \times 10^{36}$ ergs~s$^{-1}$,
are in accord with expectation from the \chandra\ results
on \asrc\ described in \S2. 
Assuming a distance of 6.6~kpc as discussed in \S4 
the ratio $L_x(2-10~{\rm keV})/\dot E = 8 \times 10^{-3}$ for the pulsar,
and $1 \times 10^{-3}$ for its PWN.  The 20$-$300~keV flux of \asrc\ measured
by \integral\ \citep[$9 \times 10^{-11}$
ergs~cm$^{-2}$~s$^{-1}$;][]{mal05} corresponds to a luminosity of
$0.09\,\dot E\,d_{6.6}^2$, and the $> 200$~GeV flux from \tev\
\citep[$3.3 \times 10^{-11}$ ergs~cm$^{-2}$~s$^{-1}$;][]{aha06}
corresponds to $0.03\,\dot E\,d_{6.6}^2$.

\subsection{\xte\ Spectral Analysis}

The spectrum of the pulsed flux from \psr\ can be isolated using
phase-resolved spectroscopy. We used the {\it fasebin} software to
construct phase-dependent spectra based on the ephemeris of
Table~\ref{tab:ephem}.  For each epoch and PCU we
constructed spectra from the top Xenon layer only
and combined them to produce single spectra per
PCU for the entire set of observations.  Similarly, standard PCA
responses for each PCU were generated at each epoch and averaged.  In
fitting the pulsed flux, the unpulsed emission provides a near perfect
background estimate.  The spectra were divided into two groups of 0.5 cycles
each to represent the ``off-peak'' and ``on-peak'' emission and
were fitted in the $2-20$~keV range using XSPEC.
The spectra from the four active PCUs were analyzed, both separately
and combined into a single spectrum, which gave similar results.

A simple absorbed power-law model was used with the interstellar
absorption held fixed at $N_{\rm H} = 4.5\times10^{22}$~cm$^{-2}$
determined from the \chandra\ fit; leaving the $N_{\rm H}$
unconstrained results in a larger uncertainty in $\Gamma$.  The
resulting best-fit photon index is $\Gamma = 1.2(1.1-1.3)$.  The
unabsorbed 2$-$10 keV pulsed flux is $9 \times
10^{-12}$~ergs~cm$^{-2}$~s$^{-1}$, which represents, within
statistics, all of the point-source flux of \psr\ measured by
\chandra\ $(8.8 \times 10^{-12}$~ergs~cm$^{-2}$~s$^{-1}$), indicating
that its intrinsic pulsed fraction is $\sim 100\%$.  The corresponding
pulsed luminosity (assumed isotropic) is $4.6 \times
10^{34}\,d_{6.6}^2$~ergs~s$^{-1}$.

The \xte\ power-law slope derived from the pulsed emission is slighlty
flatter than $\Gamma = 1.5$ reported by \citet{mal05} for the joint
fit to \asca\ and \integral\ spectra of \asrc. We also find
that this result is at odds with the \chandra\ spectra
presented in \S2.1, which indicate $\Gamma = 0.5 \pm 0.2$ for the
dominant pulsar component.  The difference is sufficiently significant
to suggest that a steepening of the spectrum of the pulsar occurs in
the 8$-$15 keV band.

\begin{figure}[t]
\centerline{
\includegraphics[height=0.95\linewidth,angle=270,clip=true]{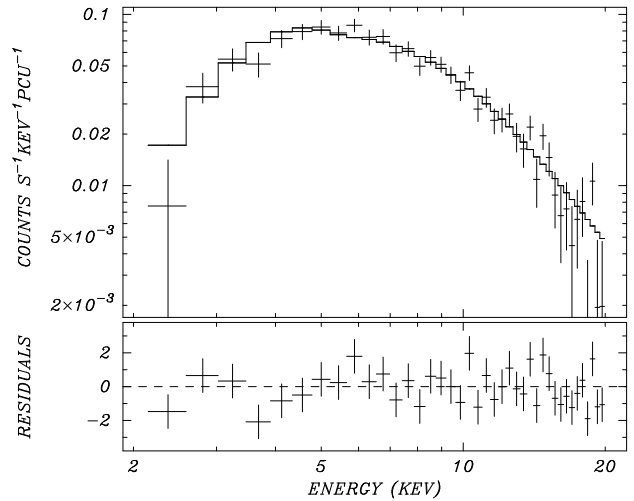}
}
\caption{\xte\ spectrum of pulsed flux from \psr\ obtained by
subtracting the off-peak spectrum from the on-peak spectrum,
and fitting to an absorbed power-law model (see text).}
\label{spectrum}
\end{figure}

\section{Discussion}

Graphic evidence of the relationship of \asrc\ to \tev\ is
provided by the \chandra\ ACIS-I image, which resolves \asrc\
into a point source plus a centrally peaked $2^{\prime}$ diameter 
nebula near one end of the major axis of the TeV source.
Interestingly, a second pulsar/PWN system is
evident in the \chandra\ image corresponding to a weaker source
previously detected as \bsrc.
Although \bsrc\ contains a point-source component $\sim 100$ times weaker
than that of \asrc, the nebular fluxes of the two sources are
virtually identical.  This is not too surprising, as the
ratio of pulsar X-ray luminosity
to spin-down power is such a widely ranging quantity \citep{kar08}.
The mere presence of the second PWN is of relevance here
because it indicates another likely energetic pulsar
of $\dot E > 4 \times 10^{36}$ ergs~s$^{-1}$ \citep{got03},
which implies that the two sources are {\it a priori}
comparably suited to powering the TeV emission.
Since \bsrc\ is located near the opposite end of the elongated
TeV source from \asrc, the possibility that \tev\ comprises
two PWNe, perhaps born in the same star cluster,
is an intriguing question for further investigation.

The location of \asrc\ near an unusual cluster of red supergiant stars
\citep[RSGC1;][]{fig06} suggests a distance of 6.6~kpc that \cite{dav08}
determined from the radial velocity of the cluster.
Davies et al. also measured
the infrared extinction to the cluster, $A_K = 2.6$ mag,
which is equivalent to $A_V = 23.5$,
or $N_{\rm H} = 4 \times 10^{22}$~cm$^{-2}$ according to the
conversion $N_{\rm H}/A_V = 1.79 \times 10^{21}$ cm$^{-2}$~mag$^{-1}$
\citep{pre95}.  Our X-ray fitted $N_{\rm H}$ is consistent with this.
It is plausible
that \psr\ was born in this young cluster, which is centered
$3^{\prime}$ north-northwest of the X-ray source (see Figure~1).
At this distance,
a 23 kyr old neutron star born in the center of the cluster
requires a transverse velocity of $245$ km~s$^{-1}$ to
reach its present location, less if it
was born in the outskirts.

It is possible that the TeV PWN is displaced in one direction from
the pulsar by an asymmetric reverse shock resulting from a supernova that
exploded initially in an inhomogeneous medium, an explanation
suggested by \citet{blo01} for the similar appearance of the Vela X remnant
\citep{aha06b}.  Other likely associations that show such displacements
are PSR B1823$-$13/HESS J1825$-$137 \citep{aha06c,pav08},
PSR J1809$-$1917/HESS J1809$-$193 \citep{aha07,kar07},
PSR J1718$-$3825/HESS J1718$-$385 \citep{aha07,hin07}, and
possibly PSR J1617$-$5055/HESS J1616$-$508 \citep{aha06,lan07}.

Based on the number of red supergiants in the cluster RSGC1,
\citet{dav08} estimate that it has a
total mass of $(2-4)\times 10^4\,M_{\odot}$,
the highest of any young cluster in the Galaxy.
\citet{fig06} calculated that it should
produce a supernova every 40$-$80~kyr on average, which is frequent
enough statistically to expect a pulsar of the 23~kyr characteristic
age that we measured for \asrc.
In fact, it is statistically suggestive that the previous pulsar
born in this cluster is \bsrc, and that it  contributes
to \tev.  Theoretically, TeV sources powered by
pulsars can live for up to 100~kyr as inverse Compton emitters
in regions of low ($\mu$G) $B$-fields \citep{dej08}.

A plausible mechanism of the TeV emission from \tev\ is
inverse Compton scattering of ambient photons off relativistic
electrons in an extended PWN, with the
target IR/optical photons supplied by the star cluster.
Considering the luminosity of only the IR detected supergiants of
RSGC1 as a lower limit on the seed photons
\citep[$\approx 2.8 \times 10^6\ L_{\odot}$:][]{fig06,dav08},
at a distance of $3 \times 10^{19}$ cm from the cluster
the energy density of optical/IR photons in \tev\ is
$\sim 20$~eV~cm$^{-3}$.  If the pulsar injected $10^{37}$ e$^{\pm}$~s$^{-1}$
into this region for the past $10^4$~yr
with average $\gamma=10^6$, then the inverse Compton
luminosity can be $\leq 2.7 \times 10^{36}$ ergs~s$^{-1}$.
This approximation does not account for the Klein-Nishina suppression
of the cross section, which would reduce it.  Nevertheless,
it appears sufficient to supply the
$1.7 \times 10^{35}\,d_{6.6}^2$ ergs~s$^{-1}$ luminosity of
\tev, as it does not even invoke the higher luminosity
from the pulsar in its younger phase, which energy may still
be present as relic electrons in the extended TeV nebula.

\section{Conclusions}

Using \chandra, we resolved the X-ray image and spectrum of \asrc\
into its pulsar and PWN components, with 2$-$10~keV luminosities of
$L_{PSR} = 4.6 \times 10^{34}\ d_{6.6}^2$~ergs~s$^{-1}$ and $L_{PWN} =
5.2 \times 10^{33}\ d_{6.6}^2$~ergs~s$^{-1}$.  Led by this evidence
we searched for pulsations using \xte\ and discovered \psr\ in
\asrc\ at a period of 70.5~ms.  Timing its spin-down, we
found $\dot E = 5.5 \times 10^{36}$ ergs~s$^{-1}$, a value that is
consistent with pulsars of similar X-ray luminosity and X-ray nebular
properties.  The 20$-$300 keV luminosity observed by \integral\
\citep{mal05} is $\approx 4.7 \times 10^{35}\,d_{6.6}^2$
ergs~s$^{-1}$, and is likely to be an extension of the hard pulsar
spectrum rather than the PWN.  The 0.2$-$20~TeV luminosity of \tev\ is
$1.7 \times 10^{35}\,d_{6.6}^2$ ergs~s$^{-1}$, which is 3\% of the
pulsar $\dot E$, while the TeV emission is several times larger in
extent than the X-ray PWN.  \psr\ is located at one end of the
elongated TeV source.

The 23~kyr characteristic age of \psr\ is comparable to those
of the middle-aged pulsars Vela and PSR B1823$-$13, with which it shares
the characteristic of an X-ray PWN displaced from the center of
a larger TeV nebula.  Such pulsars are old enough that the
displacement can be enforced by
a reverse supernova shock propagating in an inhomogeneous
interstellar medium.
Alternatively or in addition,
we may be seeing relic electrons from 
asymmetric diffusion out of the PWN in an earlier phase.

We also discovered a \chandra\ point source
inside a second diffuse X-ray source adjacent
to \tev.  Probably an older pulsar/PWN,
\bsrc\ may also contribute to \tev, explaining
its elongated shape as a blended source.
This scenario may be testable with spatially resolved
spectroscopy of the TeV emission, especially if the
age of the putative second pulsar is significantly
different from the first.
Although not yet conclusively identified as a pulsar,
\bsrc\ should have similar spin-down luminosity as \asrc.
A deep search for radio pulsations from \bsrc\ is warranted
to determine its spin-down parameters,
it being too faint for an X-ray search.

\acknowledgements

We thank the \xte\ mission for making observing time available for
this project. We are especially grateful to Drs. Jean Swank and Craig
Markwardt for their invaluable assistance in
planning these observations. This work also made use of archival data
from the \chandra\ X-ray Observatory Center. Financial support was
provided by the National Aeronautics and Space Administration through
\chandra\ Award SAO G08-9060X issued by the \chandra\ X-ray
Observatory Center, which is operated by the Smithsonian Astrophysical
Observatory for and on behalf of NASA under contract NAS8-03060.

\end{document}